\documentclass[a4paper]{article}

\usepackage[T1]{fontenc}

\usepackage{amsfonts}

\usepackage{algpseudocode} 
\usepackage{algorithm}
\usepackage{program}

\usepackage{isabelle}
\usepackage{isabellesym}
\usepackage{isabelletags}

\usepackage{hyperref}

\providecommand{\keywords}[1]{\textbf{\textit{Index terms---}} #1}

\newcommand{\linform}[1]{\mbox{\ensuremath{\textsl{L}\, {#1}}}}

\newcommand{\ima}[1]{\mbox{im {#1}}}

\newcommand{\rk}[1]{\mbox{rk {#1}}}

\newcommand{\nul}[1]{\mbox{null {#1}}}

\newcommand{\rref}[1]{\mbox{rref}\, {#1}}

\newcommand{\ncols}[1]{\mbox{ncols}\, {#1}}

\newcommand{\nrows}[1]{\mbox{nrows}\, {#1}}

\newcommand{\col}[2]{\mbox{col}\, {#1}\, {#2}}

\newcommand{\nonzero}[2]{\mbox{nonzero}\, {#1}\, {#2}}

\newcommand{\indexnonzero}[2]{\mbox{index-nonzero}\, {#1}\, {#2}}

\newcommand{\interchangerows}[3]{\mbox{interchange-rows}\, {#1}\, {#2}\, {#3}}

\newcommand{\multrow}[3]{\mbox{mult-row}\, {#1}\, {#2}\, {#3}}

\newcommand{\addrow}[4]{\mbox{add-row}\, {#1}\, {#2}\, {#3}\, {#4}}

\begin{document}

\sloppy


\title{Applications of the Gauss-Jordan algorithm, done right}

%
%
%
%
%

%

\author{Jesús Aransay%
    \\  \url{http://www.unirioja.es/cu/jearansa}
    \\  Departamento de Matemáticas y Computación,
    \\  Universidad de La Rioja, Spain
    \and
        Jose Divasón%
        \\ \url{http://www.unirioja.es/cu/jodivaso}
        \\ Departamento de Matemáticas y Computación,
        \\ Universidad de La Rioja, Spain}

\date{\today}

\maketitle

\begin{abstract}

Computer Algebra systems are widely spread because of some of their remarkable
features such as their ease of use and performance. Nonetheless, this focus on
performance sometimes leads to unwanted consequences: algorithms and computations
are implemented and carried out in a way which is sometimes not transparent to the users,
and that can lead to unexpected failures. In this paper we present a formalisation in a
proof assistant system of a \emph{naive} version of the Gauss-Jordan algorithm,
with explicit proofs of some of its applications, and additionally a process to obtain
versions of this algorithm in two different functional languages (SML and Haskell)
by means of code generation techniques from the verified algorithm. The obtained programs are
then applied to test cases, which, despite the simplicity of the original algorithm,
have shown remarkable features in comparison to some Computer Algebra systems,
such as Mathematica\textsuperscript{\textregistered} (where some of these computations
are even incorrect), or Sage (in comparison to which the generated programs show a
compelling performance). The aim of the paper is to show that, with the current technology
in Theorem Proving, formalising Linear Algebra procedures is a challenging but rewarding
task, which provides programs that can be compared in some aspects to \emph{state of the art}
procedures in Computer Algebra systems, and whose correctness is formally proved.

\end{abstract}



\keywords{Numerical Linear Algebra, Algorithm implementation, Isabelle/HOL, Code generation}

\section{Introduction}
\label{s_introduction}

Computer Algebra systems are used nowadays in very different environments and,
after years of continuos improvement, with an ever increasing level of confidence. Despite
this, these systems focus intensively on performance, and their algorithms are subject to
continuous refinements and modifications, which can sometimes derive in a loss of accuracy
and even sometimes of correctness. On the other hand, theorem provers are designed to prove
the correctness of program specifications and mathematical results. This task is far from
trivial, and it does not pay off in terms of performance but only in terms of the simplicity
and the insight of the programs one is trying to formalise. Consequently, one can be faced with
the very little appealing situation where a program has been formalised but its usefulness is,
at least, arguable.

Fortunately, and after years of continuous work, theorem proving tools have reduced
this well-known gap, and the technology they offer is being used to implement and also
to analyse state of the art algorithms and programs (see for instance~\cite{ESLANENISHSM13,ST13}).
In this work, we present an experiment to formalise a version of the Gauss-Jordan
algorithm over matrices in the theorem prover Isabelle/HOL. The algorithm computes
the reduced row echelon form of a matrix, which is then proved to be applicable
to solve standard problems in Linear Algebra, such as computing the rank of a linear form,
computing determinants and inverses, solving systems of linear equations, and
computing bases of fundamental subspaces of linear forms. These verified algorithms
are later code-generated to the functional languages SML and Haskell. The code obtained
in these languages is tested against a battery of examples. The algorithm that we
implement is neither specialised, nor obtained from a Computer Algebra system, but just a simple version of the
Gauss-Jordan algorithm. Nevertheless, the utility of our work is threefold. First,
it shows that the formalisation of Linear Algebra algorithms
in a theorem prover is feasible. Second, the code obtained in Haskell and SML, even
if it lacks of the performance of the specialised code of standard Computer Algebra
systems, was capable of computing some determinants with big integers that produced
a bug in Mathematica\textsuperscript{\textregistered}~\cite{DUPEVA13}. Finally, the
already existing infrastructure in the Isabelle/HOL Multivariate Analysis
Library allowed us to keep the ties among Linear Algebra algorithms and their
mathematical meaning or origin (a feature that is not possible in Computer Algebra systems).

The paper will be divided as follows: in Section~\ref{s_isabelle} we
introduce the Isabele/HOL theorem prover and the infrastructure in such system
that is used in our work; we distinguish among the parts which are already in the system,
and the ones that are product of our own work. In Section~\ref{s_the_gauss_jordan_algorithm_and_its_applications}
we present a version of the Gauss-Jordan algorithm over fields,
as well as the different applications of it that we have formalised in Isabelle/HOL.
In Section~\ref{s_code_generation_to_functional_languages} we present
the code generation process from the formalised Isabelle algorithm
to the running versions in SML and Haskell. Finally, in Section~\ref{s_conclusions_and_further_work}
we draw some conclusions and possible research lines that follow from our work.

The source files of the development are available from~\cite{ARDI13d}; they have been
developed under the Isabelle 2013-2 version. The previous web site also includes the SML
and Haskell code generated from the Isabelle specifications, and also the input matrices
that have been used in the benchmarks presented in Section~\ref{s_performance_of_the_generated_programs}.

\section{Isabelle}
\label{s_isabelle}

\subsection{Isabelle/HOL}
\label{ss_isabelle_hol}

Isabelle~\cite{PA90} is a generic theorem prover which has been instantiated to
support different object-logics, from which higher-order logic (or briefly,
\emph{HOL}~\cite{NIPAWE02}) is the one that offers a greatest number of facilities
to the user, some of which will be relevant to our work (such as code
generation~\ref{ss_code_generation}). The Isabelle metalogic is based on two
components: a (rather simple) type system, including non-empty types and function
types ($\alpha \to \beta$), from which the \emph{prop} type includes the propositions
accepted by the system, and a set of inference rules acting over elements of
\emph{prop} type, expressing the properties of the metalogic connectors (implication,
universal quantifier and logical equivalence). New propositions in the system are then
elements of type \emph{prop} that, by means of iterative applications of inference rules,
have been reduced to trivial propositions (the \emph{True} constant).

From the previous simple infrastructure, Isabelle/HOL introduces then some new connectors (specialised
versions of the metalogic ones for this particular logic) and additional axioms
(such as for instance the law of excluded middle). We briefly present here the features of
Isabelle/HOL in which our work relies on. The previous references offer further insight.

The HOL type system is based on non-empty types, function types
($\Rightarrow$) and type constructors $\kappa$ that can be applied to already existing
types (\emph{nat, bool}) or type variables ($\alpha, \beta$). Types can be also introduced
by enumeration (\emph{bool}) or by induction, as lists (by means of the \emph{datatype} command).
Additionally, new types can be also defined as non-empty subsets of already existing
types ($\alpha$) by means of the \emph{typedef} command; the command takes a set defined
by comprehension over a given type $\{x\mathrel{::}\alpha.\, P\, x\}$, and defines a new type $\sigma$.

Isabelle also introduces type classes in a similar fashion to Haskell; a type class
is defined by a collection of operators (over a single type variable) and
premises over them. For instance, the HOL Multivariate Analysis library has a type class \emph{field}
representing the algebraic structure. Concrete types (\emph{real}, \emph{rat}) can be proven
to be \emph{instances} of a given type class (\emph{field} in our example). Type classes
are also used to impose additional restrictions over type variables; for instance, the expression
($x\mathrel{::}\alpha\mathrel{::}\mbox{\emph{field}}$) imposes the constraint that the type
variable $\alpha$ possess the structure and properties stated in the \emph{field} type class, and
can be later replaced exclusively by types which are instances of that type class.

\subsection{HOL Multivariate Analysis}
\label{ss_hol_multivariate_analysis}

The HOL Multivariate Analysis (or \emph{HMA} for short) Library is a set of Isabelle theories which contains
a number of theoretical results in mathematical fields such as Analysis, Topology or
Linear Algebra. They are based on previous work of J.~Harrison in HOL-Light~\cite{HA13},
which includes proofs of intricate theorems (such as the Stone-Weierstrass theorem)
and has been used as a basis for appealing projects such as the formalisation of the
proof of the Kepler conjecture by T.~Hales. Among the fundamentals of the library,
one of the keys is the representation of n-dimensional vectors over a given type
($\mathbb{F}^n$, where $\mathbb{F}$ stands for a generic field, or in Isabelle
jargon a type variable $\alpha\mathrel{::}\mbox{\emph{field}}$).

The idea is to represent vectors over $\alpha$ by means of \emph{functions} from a
finite type variable $\beta::\mbox{\emph{finite}}$ to $\alpha$; for proving purposes,
this type definition is usually sufficient to support the generic structure $\mathbb{F}^n$.

The Isabelle type definition is as follows; the functions \emph{vec-nth} and \emph{vec-lambda}
are the morphisms between the abstract data type \emph{vec} and the underlying concrete data type,
functions with finite domain (the mathematical restrictions over $\alpha$ and $\beta$
are added only when required for formalisation purposes):

\begin{isabellebody}
\isanewline
\isacommand{typedef}\ \isacharparenleft\isasymalpha \isacharcomma\isasymbeta\isacharparenright\ vec\ \isacharequal\ UNIV\ \isacharcolon\isacharcolon\ \isacharparenleft\isacharparenleft\isasymbeta\isacharcolon\isacharcolon finite\isacharparenright\ \isasymRightarrow\ \isasymalpha\isacharparenright\ set
\isanewline
\ \isacommand{morphisms}\ vec\isacharminus nth\ vec\isacharminus lambda\ \isachardot\isachardot
\isanewline
\end{isabellebody}

The previous type also admits in Isabelle the shorter notation $\alpha\hat{\text{ }}\beta$.
The idea of using underlying finite types for vectors indices has great advantages,
as already pointed out by Harrison, from the formalisation point of view.
For instance, the type system enforces that operations on vectors (such as addition or multiplication)
are only performed over vectors of equal dimension, \emph{i.e.}, vectors whose
indexing types are exactly the same (this would not be the case if we were to use,
for instance, lists as vectors). Moreover, the functional flavour of operations
and properties over vectors is kept (for instance, vector addition can be
defined in a pointwise manner).

The representation of matrices is then derived in a natural way based on the one of vectors
by iterating the previous construction (matrices over a type $\alpha$ will be terms
of type $\alpha\hat{\text{ }}m\hat{\text{ }}n$, where $m$ and $n$ stand for finite type variables).

A subject that has been explored neither in the Isabelle HMA Library,
nor in HOL-Light, is the possibility to execute the previous data types and operations.
Another aspect that has not been explored in the HMA Library is
Numerical Linear Algebra. One of the novelties of our work is to establish a link between this
formalisation setting and a framework where algorithms can be represented and also executed.

\subsection{Code generation}
\label{ss_code_generation}

Another interesting feature of Isabelle/HOL is its code generation facility~\cite{HA13b}.
Its starting point are specifications (in the form of the different kinds
of definitions supported by the system) whose properties can be stated and proved,
and (formalised) rewriting rules that express properties from the original
specifications. From the previous \emph{code equations}, a \emph{shallow embedding}
from Isabelle/HOL to an abstract intermediate functional language (Mini-Haskell)
is performed. Finally, trivial transformations to the functional languages SML,
Haskell, Scala and OCaml are performed. The expressiveness of HOL (such as for
instance universal or existential quantifiers, or the Hilbert's $\epsilon$ operator)
is not that of functional programming languages, and therefore one must restrict herself to use
Isabelle ``executable'' specifications, if she aims at generating code from them (or prove
\emph{code equations} that refine non-executable specifications to executable ones).

The generated code satisfies a principle of \emph{partial correctness} by construction,
with respect to the properties that have been proved of it. This means that whenever an
expression $v$ is evaluated to some term $t$, $t = v$ is derivable in the
equational semantics of the intermediate language. See~\cite{HANI10, HA13} for further details.

\section{The Gauss-Jordan algorithm and its applications}
\label{s_the_gauss_jordan_algorithm_and_its_applications}

In a previous work~\cite{ARDI13}, we formalised the rank plus nullity
theorem of Linear Algebra. In our proof it is established that, given $V$
a finite-dimensional vector space over $\mathbb{R}$, $W$ a vector space over
$\mathbb{R}$, and $\tau \in \linform (V, W)$ (a linear form between $V$ and
$W$), $\dim (\ker (\tau)) + \dim (\ima (\tau)) = \dim (V)$ or, in other
notation, $\nul (\tau) + \rk (\tau) = \dim (V)$. We closely followed the
proof in~\cite{GO10}, as we follow here his notation. Unfortunately, having
formalised the previous result does not provide us with an algorithm computing
the dimension of the image and kernel sets of a given linear form.

As it has been proved in the HMA Library, every linear form between
finite-dimensional vector spaces over the field $\mathbb{R}$ is equivalent to
a matrix over $\mathbb{R}^{m * n}$, and therefore we can reduce the computation
of the dimensions of the range (or rank) and the kernel (or nullity)
of a linear form to the computation of the \emph{reduced row echelon form}~\cite{RO08}
(or \emph{rref}) of a matrix; the number of nonzero rows of such matrix provides
its rank, and the number of zero rows its nullity. The Gauss-Jordan algorithm
computes the \emph{rref} of a matrix.

We have formalised in Isabelle the following version of the Gauss-Jordan algorithm:

\begin{algorithm}
  \caption{Gauss-Jordan elimination algorithm}
  \begin{algorithmic}[1]
    \State {{\bf Data:} $A$ is the input matrix;}

    \Comment{$l$ is the index where the pivot is placed}
    \State $l \gets 0$;
    \For  {$k \gets 0, (\ncols A) - 1$}

        \Comment{Check that col. $k$ contains a pivot over index $l$}
        \If {$\nonzero l (\col k A)$}

            \Comment{Let $i$ be the index of first nonzero entry over $l$}
            \State $i \gets \indexnonzero l (\col k A)$

            \Comment{Rows $i$ and $l$ are interchanged}
            \State $A \gets \interchangerows A i l$

            \Comment{Row $l$ is multiplied by $(1 / A\, l\, k)$}
            \State $A\, l \gets \multrow  A  l (1 / A\, l\, k)$
            \For {$t \gets 0, (\nrows A) - 1$}
                \If {$t \ne l$}

                    \Comment{Row $t$ is added row $l$ times $(- A\, t\, k)$}
                    \State $A\, t \gets \addrow A t l (- A\, t\, k)$
                \EndIf
            \EndFor
            \State $l \gets l + 1$
        \EndIf
    \EndFor
  \end{algorithmic}
\label{A:gauss}
\end{algorithm}

The algorithm traverses the columns of the input matrix, finding in each column $k$
a pivot $i$ (the first nonzero element in a row greater than index $l$); if the pivot exists,
rows $i$ and $l$ are interchanged (if the matrix has maximum rank, $l$ will be equal to
the column index, otherwise it will be smaller), and row $l$ is multiplied by the inverse of
the pivoted element; this row is used to perform row operations to reduce all remaining coefficients
in column $k$ to $0$. If a column does not contain a pivot, the algorithm processes
the next column. The algorithm performs exclusively \emph{elementary row operations}. We have
expressed it above with \emph{imperative} constructs such as \emph{for} and variable assignments
that are not native to the Isabelle/HOL specification language. In our Isabelle specification,
rows and columns are assigned finite enumerable types, over which matrices are represented
as functions. The previous algorithm operations are expressed by means of functions
representing the output matrix after each operation. Note that a matrix
is defined by means of a function over the rows type of functions over the columns type.
For instance, the Isabelle definition shown below is the one selecting a pivot $i$ in column
$k$ over the index $l$ (line 5 in Algorithm~\ref{A:gauss}),
interchanging rows $i$ and $l$ (line 6), multiplying the row $l$ by the multiplicative
inverse of $A\, l\, k$ (line 7) and reducing the rest of the rows of the matrix,
by means of a lambda expression, which represents the new created matrix (lines 8 to 12).
The traversing operation over columns (line 3) is represented by means of a \emph{fold}
operation over the list containing the columns type universe.

\begin{isabellebody}
\isanewline
\isamarkuptrue%
Gauss{\isacharunderscore}Jordan{\isacharunderscore}in{\isacharunderscore}pos\ A\ l\ k\ {\isacharequal}\
\isanewline
\ {\isacharparenleft}let\
\isanewline
\ \ \ i\ {\isacharequal}\ {\isacharparenleft}LEAST\ n{\isachardot}\ A\ {\isachardollar}\ n\ {\isachardollar}\ k\ {\isasymnoteq}\ {\isadigit{0}}\ {\isasymand}\ n\ {\isasymge}\ l{\isacharparenright}{\isacharsemicolon}\
\isanewline
\ \ \ interchange{\isacharunderscore}A\ {\isacharequal}\ {\isacharparenleft}interchange{\isacharunderscore}rows\ A\ i\ l{\isacharparenright}{\isacharsemicolon}\
\isanewline
\ \ \ A{\isacharprime}\ {\isacharequal}\ mult{\isacharunderscore}row\ interchange{\isacharunderscore}A\ l\ {\isacharparenleft}{\isadigit{1}}{\isacharslash}interchange{\isacharunderscore}A{\isachardollar}l{\isachardollar}k{\isacharparenright}\
\isanewline
\ \ \ \ in\
\isanewline
\ \ vec{\isacharunderscore}lambda{\isacharparenleft}{\isasymlambda}t{\isachardot}\ if\ t{\isacharequal}l\ then\ A{\isacharprime}{\isachardollar}l\
\isanewline
\ \ \ \ else\ {\isacharparenleft}row{\isacharunderscore}add\ A{\isacharprime}\ t\ l\ {\isacharparenleft}{\isacharminus}{\isacharparenleft}interchange{\isacharunderscore}A{\isachardollar}t{\isachardollar}k{\isacharparenright}{\isacharparenright}{\isacharparenright}{\isachardollar}t{\isacharparenright}{\isacharparenright}%
\isanewline
\end{isabellebody}

The algorithm has several variants, both to speed up its performance and also to avoid numerical stability
issues with floating point numbers~\cite[Ch. 9]{GO10}, but in order to reduce the complexity
of its formalisation we chose the presented one. As we show later
(Section~\ref{s_performance_of_the_generated_programs}) its performance is noticeable.

The rref of a matrix has indeed further applications than computing the rank; based on the fact that
the version of Gauss-Jordan used to obtain it is based on elementary row operations, it can be also used
for the following ends:
\begin{itemize}
\item Computation of the inverse of a matrix, by ``storing'' the elementary row operations over the identity matrix.
\item Determinants, taking into account that some of the elementary row operations can introduce multiplicative constants.
\item Computation of bases and dimensions of the null (defined as $\{x \in  \mathbb{R}^m \mid A * x = 0\}$),
    left null ($\{x \in \mathbb{R}^n \mid x^T * A = 0\}$),
    column ($\{A * x \mid x \in \mathbb{R}^m\}$)
    and row ($\{A^T * x \mid x \in \mathbb{R}^n\}$) subspaces of a matrix.
\item Solutions of systems of linear equations ($\{x \mid A * x = b \wedge x \in \mathbb{R}^m\}$),
    both consistent (with unique or multiple solutions) and inconsistent ones.
\end{itemize}

The formalisation of the Gauss-Jordan algorithm and the different applications
that are presented above summed up $8\,000$ lines of code; the proofs are
devoted to check that the defined objects (determinant, inverse matrix,
solution of the linear system) are preserved (or modified in a certain way)
after each algorithm step (and more concretely, after each row operation).
By using product types, we store the input matrix and we set an initial
value for the defined object. In the case of determinants, the initial
pair is $(1, A)$. The other computations start from $(I_{n}, A)$ or
$(I_{m}, A^T)$. After each algorithm step, the corresponding modification
is applied to the first component. In the computation of each of the previous
pairs, there is a notion of \emph{invariant} that is preserved through the
Gauss-Jordan algorithm steps. For instance, in the case of determinants,
given a matrix $A$, after $n$ elementary operations the pair $(b_n, A_{n})$
is obtained, and it holds that $\det A = b_n * (\det A_n)$. Since the
algorithm is terminating (the elements indexing the columns are an
enumerable type), after a finite number $m$ of operations we obtain a pair $(b_m, \rref A)$
such that $\det A = b_m * (\det (\rref A))$; since we proved that the determinant
of $\rref A$ is the product of its diagonal elements, the computation is completed.

In a similar way we perform the proof of the computation of the inverse of a matrix
(starting from an input square matrix $A$ of dimension $n$, the pair $(I_n, A)$
is built and after every row operation, $(P', A')$ is such that $P' * A = A'$),
as long as $A$ is invertible (in other words, $\rref A = I_n$). When
the Gauss-Jordan algorithm reaches $\rref A$, the first component of the pair
holds the matrix $P$, which is the product of every elementary
operation performed. The computation of the bases of the fundamental
subspaces of linear forms are also based on the computation of the
matrix $P$ generated from applying the Gauss-Jordan algorithm to $A$
(or $A^T$) and the same operations to $I_n$ ($I_m$). Their Isabelle definitions follow:

\begin{isabellebody}
\isanewline
\isacommand{definition}\isamarkupfalse%
\ basis{\isacharunderscore}null{\isacharunderscore}space\ A\ {\isacharequal}\
\isanewline
{\isacharbraceleft}row\ i\ {\isacharparenleft}P{\isacharunderscore}Gauss{\isacharunderscore}Jordan\ {\isacharparenleft}transpose\ A{\isacharparenright}{\isacharparenright}
\isanewline
\ \ {\isacharbar}i{\isachardot}\ to{\isacharunderscore}nat\ i\ {\isasymge}\ rank\ A{\isacharbraceright}
\isanewline
\isacommand{definition}\isamarkupfalse%
\ basis{\isacharunderscore}row{\isacharunderscore}space\ A\ {\isacharequal}\
{\isacharbraceleft}row\ i\ {\isacharparenleft}Gauss{\isacharunderscore}Jordan\ A{\isacharparenright}\
\isanewline\ \ {\isacharbar}i{\isachardot}\ row\ i\ {\isacharparenleft}Gauss{\isacharunderscore}Jordan\ A{\isacharparenright}\ {\isasymnoteq}\ {\isadigit{0}}{\isacharbraceright}
\isanewline
\isacommand{definition}\isamarkupfalse%
\ basis{\isacharunderscore}col{\isacharunderscore}space\ A\ {\isacharequal}\
\isanewline {\isacharbraceleft}row\ i\ {\isacharparenleft}Gauss{\isacharunderscore}Jordan\ {\isacharparenleft}transpose\ A{\isacharparenright}{\isacharparenright}\
\isanewline
\ \ {\isacharbar}i{\isachardot}\ row\ i\ {\isacharparenleft}Gauss{\isacharunderscore}Jordan\ {\isacharparenleft}transpose\ A{\isacharparenright}{\isacharparenright}\ {\isasymnoteq}\ {\isadigit{0}}{\isacharbraceright}
\isanewline
\isacommand{definition}\isamarkupfalse%
\ basis{\isacharunderscore}left{\isacharunderscore}null{\isacharunderscore}space\ A\ {\isacharequal}\
\isanewline
\ {\isacharbraceleft}row\ i\ {\isacharparenleft}P{\isacharunderscore}Gauss{\isacharunderscore}Jordan\ A{\isacharparenright}\ {\isacharbar}\ i{\isachardot}\ to{\isacharunderscore}nat\ i\ {\isasymge}\ rank\ A{\isacharbraceright}%
\isanewline
\end{isabellebody}

With respect to the solution of systems of linear equations $A * x = b$,
we prove that, if a system is consistent, its set of solutions is equal to
a single point plus any element which is solution to the homogeneous system
associated to the input system of equations, $A * x = 0$ (or, in other words,
the null space of $A$). We also prove that every solution of the system must
be of this form. Therefore, in order to solve a system, we start from the pair
$(I_{n}, A)$ and after applying the Gauss-Jordan algorithm to the second component,
and the same elementary operations to the first component, $(P, \rref A)$ is
obtained. The vector $b$ is multiplied by the matrix $P$,
and from its number of nonzero positions and the rank of $A$ (or $\rref A$)
the system is classified as consistent or inconsistent. In the first case,
a single solution is computed by taking advantage of $\rref A$. The basis of
the null space is computed applying Gauss-Jordan elimination to $A^T$
in $(I_m, A^T)$, and performing similar row operations to $I_m$.

In order to consider inconsistent systems suitably, we have represented the solutions
as elements of the Isabelle option type (\emph{(SOME x. P x)}, \emph{NONE}),
which are presented as a singular point (whenever the system has solution),
and the corresponding vectors forming a basis of the null space (or the empty set).
We have formalised in Isabelle that every solution to a given system is of the previous form
(most Computer Algebra systems, and previous formalisations of the solution of systems
of linear equations through Gauss-Jordan algorithm~\cite{NI11} compute single solutions
and sometimes for exclusively compatible systems with equal number of equations and unknowns).

Regarding the complexity, Gauss-Jordan algorithm is well-known to
perform $O (n^3)$ operations for input square matrices of dimension $n * n$.
The amount of operations involved in the computation of the rref,
the rank, the determinant and the row and column spaces, following
our ideas, will be of such order (except for the computation of the transpose).
In order to compute the inverse of a matrix, the number of columns double,
but the number of rows is preserved; the arithmetic operations are twice
the number of operations performed in the rref. The same number of
operations are performed for the computation of the null and left
null spaces, since they require computing the $P$ matrix associated to $\rref A$.
Finally, computing the solutions of a system of linear equations involves
the computation of the rref of $A$ and its $P$ matrix, and also the
computation of the rref of $A^T$ and its corresponding $P$ matrix.

\section{Code generation to functional languages}
\label{s_code_generation_to_functional_languages}

The previous version of the Gauss-Jordan algorithm can be directly executed \emph{inside}
of Isabelle (by rewriting specifications and code equations) with some setup modifications
that we presented in a previous work~\cite[Sect. 4]{ARDI13b}. The specifications are themselves
executable, since we are dealing with finite types for representing matrices columns and rows.
For instance, the Isabelle function \emph{Gauss-Jordan-in-pos} presented above, that makes
use of the \emph{LEAST} operator (based itself on the Hilbert's $\epsilon$ operator),
which is not executable in general, takes advantage of the fact that its underlying
type is enumerable, and can be executed to select a pivot. Unfortunately, the
performance obtained makes the algorithm unusable in practice, except for
testing small examples. Matrices represented as functions over finite domains are reportedly
impractical. More concretely, there are two sources of inefficiency in the results obtained.
First, Isabelle is not designed as a programming language, and execution inside of the system
offers a not remarkable performance. In Section~\ref{ss_code_generation_and_serialisations}
we present a solution to \emph{translate} our specifications to
functional programming languages. Second, the data structures (functions)
that helped us to prove the correctness of the Gauss-Jordan algorithm and
its applications are optimal for formalisation, but not for execution.
Section~\ref{ss_data_type_refinements} describes a \emph{verified refinement}
between the type used for representing matrices in our formalisation (\emph{vec}
and its iterated construction) and \emph{immutable arrays}, a common data structure
in functional programming.

\subsection{Code generation and serialisations}
\label{ss_code_generation_and_serialisations}

The first problem is solved by \emph{generating} our specifications to a
programming (functional) language, as introduced in Section~\ref{ss_code_generation}.
Our choices (from the available languages in the standard Isabelle code generation
setup) were SML (since the SML Standard Library includes a \emph{Vector} type
representing immutable arrays) and Haskell (for a similar reason, with the Haskell
\emph{IArray} class type and its corresponding instance \emph{IArray.Array}, and
also because it has a type \emph{Rational} representing arbitrary precision
rational numbers).

Additionally, we make use of \emph{serialisations}, a process to map Isabelle
types and operations to corresponding ones in the target languages. Serialisations
are common practice in code generation processes (see~\cite{HA13b} for some
introductory examples); otherwise, the source types and operations would be
\emph{generated} from scratch in the target languages,
and the obtained code would be less usable and efficient (for instance,
$nat$ type would be generated to an \emph{ad-hoc} type with $0$ and $Suc$
as constructors, and then $int$ as the equivalence classes of pairs of naturals).
The following Isabelle code snippet presents the serialisation that we produced from the Isabelle
type \emph{rat} representing rational numbers (which is indeed based on equivalence classes),
to the Haskell type \emph{Rational}. As it can be observed, it identifies operations
(including type constructors) from the source and the target languages.

\begin{isabellebody}
\isanewline
\isacommand{code{\isacharunderscore}printing}\isamarkupfalse%
\isanewline
\ \ \isakeyword{type{\isacharunderscore}constructor}\ rat\ {\isasymrightharpoonup}\ {\isacharparenleft}Haskell{\isacharparenright}\ {\isachardoublequoteopen}Prelude{\isachardot}Rational{\isachardoublequoteclose}\isanewline
\ \ {\isacharbar}\ \isakeyword{class{\isacharunderscore}instance}\ rat\ {\isacharcolon}{\isacharcolon}\ {\isachardoublequoteopen}HOL{\isachardot}equal{\isachardoublequoteclose}\ {\isacharequal}{\isachargreater}\ {\isacharparenleft}Haskell{\isacharparenright}\ {\isacharminus}\ \isanewline
\ \ {\isacharbar}\ \isakeyword{constant}\ {\isachardoublequoteopen}{\isadigit{0}}\ {\isacharcolon}{\isacharcolon}\ rat{\isachardoublequoteclose}\ {\isasymrightharpoonup}\
\isanewline
\ \ \ \ {\isacharparenleft}Haskell{\isacharparenright}\ {\isachardoublequoteopen}Prelude{\isachardot}toRational\ {\isacharparenleft}{\isadigit{0}}{\isacharcolon}{\isacharcolon}Integer{\isacharparenright}{\isachardoublequoteclose}\isanewline
\ \ {\isacharbar}\ \isakeyword{constant}\ {\isachardoublequoteopen}{\isadigit{1}}\ {\isacharcolon}{\isacharcolon}\ rat{\isachardoublequoteclose}\ {\isasymrightharpoonup}\
\isanewline
\ \ \ \ {\isacharparenleft}Haskell{\isacharparenright}\ {\isachardoublequoteopen}Prelude{\isachardot}toRational\ {\isacharparenleft}{\isadigit{1}}{\isacharcolon}{\isacharcolon}Integer{\isacharparenright}{\isachardoublequoteclose}
\isanewline
\ \ {\isacharbar}\ \isakeyword{constant}\ {\isachardoublequoteopen}Frct{\isachardoublequoteclose}\ {\isasymrightharpoonup}
\isanewline
\ \ \ \ {\isacharparenleft}Haskell{\isacharparenright}\ {\isachardoublequoteopen}{\isacharparenleft}let\ {\isacharparenleft}x{\isacharcomma}y{\isacharparenright}\ {\isacharequal}\ {\isacharunderscore}\ in\ {\isacharparenleft}Rational{\isachardot}fract\
\isanewline
\ \ \ \ \ {\isacharparenleft}integer{\isacharprime}{\isacharunderscore}of{\isacharprime}{\isacharunderscore}int\ x{\isacharparenright}\ {\isacharparenleft}integer{\isacharprime}{\isacharunderscore}of{\isacharprime}{\isacharunderscore}int\ y{\isacharparenright}{\isacharparenright}{\isacharparenright}{\isachardoublequoteclose}\isanewline
\ \ {\isacharbar}\ \isakeyword{constant}\ {\isachardoublequoteopen}quotient{\isacharunderscore}of{\isachardoublequoteclose}\ {\isasymrightharpoonup}\
\isanewline
\ \ \ \ {\isacharparenleft}Haskell{\isacharparenright}\ {\isachardoublequoteopen}{\isacharparenleft}let\ x\ {\isacharequal}\ {\isacharunderscore}\ in\
\isanewline
\ \ \ \ \ {\isacharparenleft}Int{\isacharprime}{\isacharunderscore}of{\isacharprime}{\isacharunderscore}integer\ {\isacharparenleft}Rational{\isachardot}numerator\ x{\isacharparenright}{\isacharcomma}\
\isanewline
\ \ \ \ \ \ Int{\isacharprime}{\isacharunderscore}of{\isacharprime}{\isacharunderscore}integer\ {\isacharparenleft}Rational{\isachardot}denominator\ x{\isacharparenright}{\isacharparenright}{\isacharparenright}{\isachardoublequoteclose}
\isanewline
\ \ {\isacharbar}\ \isakeyword{constant}\ {\isachardoublequoteopen}HOL{\isachardot}equal\ {\isacharcolon}{\isacharcolon}\ rat\ {\isasymRightarrow}\ rat\ {\isasymRightarrow}\ bool{\isachardoublequoteclose}\ {\isasymrightharpoonup}\isanewline
\ \ \ \ {\isacharparenleft}Haskell{\isacharparenright}\ {\isachardoublequoteopen}{\isacharparenleft}{\isacharunderscore}{\isacharparenright}\ {\isacharequal}{\isacharequal}\ {\isacharparenleft}{\isacharunderscore}{\isacharparenright}{\isachardoublequoteclose}\isanewline
\ \ {\isacharbar}\ \ \isakeyword{constant}\ {\isachardoublequoteopen}op\ {\isacharless}\ {\isacharcolon}{\isacharcolon}\ rat\ {\isacharequal}{\isachargreater}\ rat\ {\isacharequal}{\isachargreater}\ bool{\isachardoublequoteclose}\ {\isasymrightharpoonup}\isanewline
\ \ \ \ {\isacharparenleft}Haskell{\isacharparenright}\ {\isachardoublequoteopen}{\isacharunderscore}\ {\isacharless}\ {\isacharunderscore}{\isachardoublequoteclose}\isanewline
\ \ {\isacharbar}\ \ \isakeyword{constant}\ {\isachardoublequoteopen}op\ {\isasymle}\ {\isacharcolon}{\isacharcolon}\ rat\ {\isacharequal}{\isachargreater}\ rat\ {\isacharequal}{\isachargreater}\ bool{\isachardoublequoteclose}\ {\isasymrightharpoonup}\isanewline
\ \ \ \ {\isacharparenleft}Haskell{\isacharparenright}\ {\isachardoublequoteopen}{\isacharunderscore}\ {\isacharless}{\isacharequal}\ {\isacharunderscore}{\isachardoublequoteclose}\isanewline
\ \ {\isacharbar}\ \isakeyword{constant}\ {\isachardoublequoteopen}op\ {\isacharplus}\ {\isacharcolon}{\isacharcolon}\ rat\ {\isasymRightarrow}\ rat\ {\isasymRightarrow}\ rat{\isachardoublequoteclose}\ {\isasymrightharpoonup}\isanewline
\ \ \ \ {\isacharparenleft}Haskell{\isacharparenright}\ {\isachardoublequoteopen}{\isacharparenleft}{\isacharunderscore}{\isacharparenright}\ {\isacharplus}\ {\isacharparenleft}{\isacharunderscore}{\isacharparenright}{\isachardoublequoteclose}\isanewline
\ \ {\isacharbar}\ \isakeyword{constant}\ {\isachardoublequoteopen}op\ {\isacharminus}\ {\isacharcolon}{\isacharcolon}\ rat\ {\isasymRightarrow}\ rat\ {\isasymRightarrow}\ rat{\isachardoublequoteclose}\ {\isasymrightharpoonup}\isanewline
\ \ \ \ {\isacharparenleft}Haskell{\isacharparenright}\ {\isachardoublequoteopen}{\isacharparenleft}{\isacharunderscore}{\isacharparenright}\ {\isacharminus}\ {\isacharparenleft}{\isacharunderscore}{\isacharparenright}{\isachardoublequoteclose}\isanewline
\ \ {\isacharbar}\ \isakeyword{constant}\ {\isachardoublequoteopen}op\ {\isacharasterisk}\ {\isacharcolon}{\isacharcolon}\ rat\ {\isasymRightarrow}\ rat\ {\isasymRightarrow}\ rat{\isachardoublequoteclose}\ {\isasymrightharpoonup}\isanewline
\ \ \ \ {\isacharparenleft}Haskell{\isacharparenright}\ {\isachardoublequoteopen}{\isacharparenleft}{\isacharunderscore}{\isacharparenright}\ {\isacharasterisk}\ {\isacharparenleft}{\isacharunderscore}{\isacharparenright}{\isachardoublequoteclose}\isanewline
\ \ {\isacharbar}\ \isakeyword{constant}\ {\isachardoublequoteopen}op\ {\isacharslash}\ {\isacharcolon}{\isacharcolon}\ rat\ {\isasymRightarrow}\ rat\ {\isasymRightarrow}\ rat{\isachardoublequoteclose}\ {\isasymrightharpoonup}\isanewline
\ \ \ \ {\isacharparenleft}Haskell{\isacharparenright}\ {\isachardoublequoteopen}\ {\isacharparenleft}{\isacharunderscore}{\isacharparenright}\ {\isacharprime}{\isacharslash}\ {\isacharparenleft}{\isacharunderscore}{\isacharparenright}{\isachardoublequoteclose}\isanewline
\ \ {\isacharbar}\ \isakeyword{constant}\ {\isachardoublequoteopen}uminus\ {\isacharcolon}{\isacharcolon}\ rat\ {\isacharequal}{\isachargreater}\ rat{\isachardoublequoteclose}\ {\isasymrightharpoonup}\isanewline
\ \ \ \ {\isacharparenleft}Haskell{\isacharparenright}\ {\isachardoublequoteopen}Prelude{\isachardot}negate{\isachardoublequoteclose}
\isanewline
\end{isabellebody}

The complete set of Isabelle serialisations that we have taken advantage of are
shown in Table~\ref{T_serialisations}. The Isabelle types \emph{rat}, \emph{real} and \emph{bit}
represent respectively $\mathbb{Q}$, $\mathbb{R}$ and $\mathbb{Z}_2$. The SML type
\emph{IntInf.int} represents arbitrary precision integers. It is worth noting that
the Isabelle type \emph{real} can be also serialised to the ones used for \emph{rat}
in SML and Haskell, preserving arbitrary precision and avoiding numerical stability
issues. Types presented in bold face identify serialisations that were introduced by
us as part of this work. We also contributed some improvements to the Isabelle Library
in the serialisation to the SML type \emph{Vector}.

\begin{table}
\centering
\caption{Type serialisations}
\begin{tabular}{|c|c|c|}
\hline
Isabelle/HOL     & SML                            & Haskell                    \\ \hline
\emph{iarray}    & \emph{Vector}                  & {\bf \emph{IArray.Array}}  \\ \hline
\emph{rat}       & \emph{IntInf.int / IntInf.int} & {\bf \emph{Rational}}      \\ \hline
\emph{real}      & \emph{Real.real}               & {\bf \emph{Double}}        \\ \hline
\emph{bit}       & {\bf \emph{IntInf.int}}        & {\bf \emph{Integer}}       \\
\hline
\end{tabular}
\label{T_serialisations}
\end{table}

The SML Standard Library lacks of a type representing arbitrary
precision rational numbers, and thus the proposed serialisation for
\emph{rat} is quotients of arbitrary precision integers. As we will
see in our performance tests (see~Section~\ref{s_performance_of_the_generated_programs})
Haskell will take advantage of its native \emph{Rational} type to outperform SML.
We explored the possibilities of double-precision floating-point formats
(\emph{Double} in Haskell, \emph{Real.real} in SML) in the target languages
in the search for a wider comparison of our algorithm
with Computer Algebra systems. The \emph{bit} type admits multiple
serialisations, ranging from boolean values to subsets of the integers. Experimental
results showed us that the better performing option was to serialise
\emph{bit} and its operations to integers in the target language
and operations \emph{modulo} 2.

\subsection{Data type refinements}
\label{ss_data_type_refinements}

Some data types present better properties for specification and formalisation
purposes. For instance, specifying an algorithm over sets is easier than doing
so over lists. However, the latter data type is better suited for execution tests.
Following this idea, the poor performance presented by functions representing
matrices can be solved by means of a \emph{data refinement} to a better performing data structure.

Data refinement~\cite{HAKRKUNI13} offers the possibility to replace an abstract data
type in an algorithm by a concrete one; more concretely, our intention is to replace
the \emph{vec} type representing vectors by means of a better performing type
in the code generation process. In our development, we have used the
Isabelle type \emph{iarray} as the target type of our refinement. Accordingly,
we define functions \emph{vec-to-iarray} (and \emph{matrix-to-iarray}) that convert
elements of type \emph{vec} to elements of type \emph{iarray}.

\begin{isabellebody}
\isanewline
\isacommand{definition}\ vec{\isacharunderscore}to{\isacharunderscore}iarray\isacharcolon\isacharcolon{\isacharprime}a{\isacharcircum}{\isacharprime}n\isacharcolon{\isacharcolon}{\isacharbraceleft}mod{\isacharunderscore}type{\isacharbraceright}
\isasymRightarrow\ {\isacharprime}a\ iarray
\isanewline
\ \isacommand{where}\ vec{\isacharunderscore}to{\isacharunderscore}iarray\ A {\isacharequal}\
\isanewline
\ \ IArray.of{\isacharunderscore}fun\ {\isacharparenleft}{\isasymlambda}i{\isachardot}\ A{\isachardollar}{\isacharparenleft}from{\isacharunderscore}nat\ i{\isacharparenright}{\isacharparenright}
{\isacharparenleft}CARD{\isacharparenleft}{\isacharprime}n{\isacharparenright}{\isacharparenright}
\isanewline
\end{isabellebody}

Each function over elements of type \emph{vec} needs to be replaced by a new function
over type \emph{iarray}. This requires first specifying a function over the type \emph{iarray},
and then proving that it behaves as the one over type \emph{vec}.

\begin{isabellebody}
\isanewline
\isacommand{lemma}\isamarkupfalse\ {\isacharbrackleft}code{\isacharunderscore}unfold{\isacharbrackright}{\isacharcolon}\isanewline
\ \ \isakeyword{shows}\ matrix{\isacharunderscore}to{\isacharunderscore}iarray\ {\isacharparenleft}Gauss{\isacharunderscore}Jordan\ A{\isacharparenright}\ {\isacharequal}\
\isanewline
\ \ \ Gauss{\isacharunderscore}Jordan{\isacharunderscore}iarrays\ {\isacharparenleft}matrix{\isacharunderscore}to{\isacharunderscore}iarray\ A{\isacharparenright}
\isanewline
\end{isabellebody}

The previous lemma certifies that replacing the function \emph{Gauss-Jordan} (defined
over abstract matrices, or elements of type \emph{vec}) by \emph{Gauss-Jordan-iarrays}
is correct. As it can be observed, the lemma does not include premises; lemmas
including premises cannot be used to get code generation, since premises
could not be \emph{checked} in the target languages. The label \emph{code-unfold} instructs
the code generation tool to record the lemma as a rewriting rule, replacing
occurrences of the left-hand side in the execution and code generation processes
by the right-hand side. From a more general perspective, the function \emph{matrix-to-iarray}
has to be proved to be a homomorphism between the original and the refined type.

The proving effort (in lines of code) to complete this task is almost
as challenging as the one devoted to complete the formalisation of the original
algorithm ($6\,000$ code lines). In our case, we preserved the original algorithm
(we simply replace operations over the abstract type by
equivalent ones over the concrete one) but Isabelle code generator leaves
the door open to algorithmic refinements (obviously, when the differences between
the original and the final algorithms are greater, the proving effort
to fill such a gap will be more intense).

\section{Performance of the generated programs}
\label{s_performance_of_the_generated_programs}

Once the previous serialisations have been completed, and going through the
data type refinements presented in Section~\ref{s_code_generation_to_functional_languages},
our original specification of the Gauss-Jordan algorithm and the different
applications presented in Section~\ref{s_the_gauss_jordan_algorithm_and_its_applications}
are code generated to both SML and Haskell. The automatically generated code sums
up $2\,500$ lines in SML and $2\,400$ in Haskell.

For the execution tests, we use the Poly/ML interpreter (version 5.5.1),
the MLton optimizer compiler (version 20100608), and also the Haskell compiler
GHC (version 7.4.1). We also include the Sage Mathematical Software System (5.13)
in the comparison to establish a link between our performance results and the ones
of a ``real'' system, even if we have not explored the algorithms implemented in Sage.
In the tests about determinants, we also comment on a bug in
Mathematica\textsuperscript{\textregistered} 8.0, 9.0 and 9.0.1.

The tests have been carried out in a personal computer with an
Intel\textsuperscript{\textregistered} Icore\texttrademark i5-3360M processor
(up to 2.8 GHz, 2 cores with 4 threads) with 4GB RAM memory.

Some preliminary experiments had been already carried out
in MLton and Poly/ML, exclusively for the computation of the rank of
linear forms~\cite{ARDI13c}, but developers of both tools suggested
us improvements in our methodology that eliminated the \emph{processing}
time of the input matrices (which in MLton showed to be the real
bottleneck, and also in Poly/ML a great waste of time, see the figures
in~\cite{ARDI13c}). In our first experiments, the input matrices,
of size up to $2\,560 * 2\,560$ were directly introduced in the system by means
of an explicit binder, using the SML \emph{val} command, as static data. At least in MLton,
an intermediate type checker was getting extremely slow with input data of
considerable size. The Poly/ML maintainer also modified the system behaviour in the
SVN version of the tool (and now in the 5.5.1 stable version) to improve
the processing capabilities of big inputs. From our side, we changed our
methodology to input matrices from external files by means of an ad-hoc parser.
Processing input matrices this way (at least, up to sizes of $2\,560 * 2\,560$)
proved to be no time consuming.

We present here a fragment of the experiments carried out with the new
methodology. We completed experiments of the different applications of the
Gauss-Jordan algorithm presented in Section~\ref{s_the_gauss_jordan_algorithm_and_its_applications}
in the fields $\mathbb{Z}_2$, $\mathbb{Q}$ and $\mathbb{R}$.

Table~\ref{T_rref_z2} presents the results of computing the rref of $\mathbb{Z}_2$ matrices.
As it can be noticed, Sage greatly outperforms our programs. The computing times of our programs
grow linearly compared to the number of elements in the input matrices. It is noticeable that
Poly/ML, which is an interpreter, performs better than an optimiser compiler as MLton. Haskell
seems to run poorly when the number of elements grows.

\begin{table}
\centering
\begin{tabular}{| c | c | c | c | c |}
\hline
\multicolumn{1}{|c|}{{\bf Size (n)}} & \multicolumn{1}{c|}{{\bf Poly/ML}} & \multicolumn{1}{c|}{{\bf MLton}}
& \multicolumn{1}{c|}{{\bf Haskell}} & \multicolumn{1}{c|}{{\bf Sage}} \\ \hline
100 & 0.04 & 0.06 & 6.26 & 0.04\\
200 & 0.25 & 0.46 & 49.24 & 0.04\\
300 & 0.85 & 1.52 & 170.39 & 0.04\\
400 & 2.01 & 3.52 & - & 0.04\\
500 & 3.90 & 6.87 & - & 0.04\\
600 & 6.16 & 11.77 & - & 0.04\\
800 & 15.96 & 27.98 & - & 0.04\\
1\,000 & 32.08 & 54.65 & - & 0.04\\
1\,200 & 62.33 & 94.25 & - & 0.05\\
1\,400 & 97.16 & 152.06 & - & 0.05\\
1\,600 & 139.70 & 225.76 & - & 0.05\\
1\,800 & 203.10 & 323.84 & - & 0.05\\
2\,000 & 284.28 & 437.35 & - & 0.05\\
\hline
\end{tabular}
\caption{Elapsed time (in seconds) to compute the rref of randomly generated $\mathbb{Z}_{2}^{n\times n}$ matrices.}
\label{T_rref_z2}
\end{table}

Interestingly, the rank (the number of nonzero rows of the rref) of
$\mathbb{Z}_2$ matrices permits the computation of the number of connected components
of a digital image~\cite{HEDEMAMOPOSI12}; in Neurobiology, this technique can be used
to compute the number of synapses in a neuron (see~\cite{ARDI13c}). With our programs,
the computations can be carried out on images of $2560 * 2560$ px. (which are conventional sizes
in real life experiments). The algorithm performs better with these matrices than with
randomly generated ones (as the ones used in these tests).

Table~\ref{T_det_Q} presents the performance tests to compute determinants of matrices
over $\mathbb{Q}$. Apparently, Haskell takes advantage of its native \emph{Rational} type
to get better results than Poly/ML and MLton. Sage seems to require more time than
for the rank computations with elements of type $\mathbb{Z}_2$.

\begin{table}
\centering
\begin{tabular}{| c | c | c | c | c |}
\hline
\multicolumn{1}{|c|}{{\bf Size (n)}} & \multicolumn{1}{c|}{{\bf Poly/ML}} & \multicolumn{1}{c|}{{\bf MLton}}
& \multicolumn{1}{c|}{{\bf Haskell}} & \multicolumn{1}{c|}{{\bf Sage}} \\ \hline
 10 &   0.02 &   0.01 &  0.06 & 0.00 \\
 20 &   0.35 &   0.10 &  0.12 & 0.00 \\
 30 &   1.92 &   0.61 &  0.51 & 0.00 \\
 40 &   6.70 &   2.17 &  1.60 & 0.01 \\
 50 &  20.26 &   6.73 &  4.34 & 0.01 \\
 60 &  43.02 &  14.38 &  8.86 & 0.02 \\
 70 &  87.20 &  29.20 & 17.16 & 0.03 \\
 80 & 155.14 &  51.56 & 29.81 & 0.04 \\
 90 & 263.60 &  88.15 & 49.22 & 0.05 \\
100 & 425.75 & 142.38 & 74.23 & 0.11 \\
\hline
\end{tabular}
\caption{Elapsed time (in seconds) to compute the determinant of randomly generated $\mathbb{Q}$ matrices.}
\label{T_det_Q}
\end{table}

The case of determinants of rational (or integer) matrices is specially interesting.
Varona \emph{et al}~\cite{DUPEVA13} detected that Mathematica\textsuperscript{\textregistered},
in its versions 8.0, 9.0 and 9.0.1, was computing erroneously determinants of matrices of big integers,
even for small dimensions (in their work they present an example of
a matrix of dimension $14 * 14$). The situation is such that even the same determinant,
computed twice, produces two different results. The bug was reported to the
Mathematica\textsuperscript{\textregistered} support service. The error might
be originated in the use of some arithmetic operations module large primes
(that could be either not large enough, or not as many as required).
With our verified program, the computation takes \emph{ca.} 5 seconds (in Haskell;
Poly/ML and MLton are slower, but also reach the same result), and the result obtained
is the same as in Sage (which again requires 0.0 seconds) and Maple\texttrademark.
Our algorithm relies on the arbitrary precision integer numbers used in each one
of the functional languages used, but does not contain further optimisations. The
computing time in Mathematica\textsuperscript{\textregistered} (of the wrong result)
sums up 4.32 seconds. It is worth recalling that the code of our programs
in SML and Haskell has been generated from a verified algorithm.

Table~\ref{T_inv_R} presents the times used to compute the inverse of matrices
of elements in $\mathbb{R}$. Note that here both SML and Haskell are using double-precision
floating-point numbers and thus numerical stability problems arise (the rref
of matrices is not diagonal anymore, it contains small nonzero entries), as also happens in Sage.
Once again, Poly/ML and MLton outperform Haskell. The comparison with Sage is better than in
the previous cases.

\begin{table}
\centering
\begin{tabular}{| c | c | c | c | c |}
\hline
\multicolumn{1}{|c|}{{\bf Size (n)}} & \multicolumn{1}{c|}{{\bf Poly/ML}} & \multicolumn{1}{c|}{{\bf MLton}}
& \multicolumn{1}{c|}{{\bf Haskell}} & \multicolumn{1}{c|}{{\bf Sage}} \\ \hline
100 &  0.08 &  0.09 & 10.36 & 0.05 \\
200 &  0.57 &  0.65 & 82.10 & 0.05 \\
300 &  1.80 &  2.11 &     - & 0.06 \\
400 &  4.63 &  4.92 &     - & 0.08 \\
500 &  8.24 &  9.62 &     - & 0.11 \\
600 & 15.92 & 16.51 &     - & 0.15 \\
700 & 27.35 & 25.99 &     - & 0.20 \\
800 & 42.57 & 39.37 &     - & 0.28 \\
\hline
\end{tabular}
\caption{Elapsed time (in seconds) to compute the inverse of randomly generated $\mathbb{R}^{n \times n}$ matrices.}
\label{T_inv_R}
\end{table}

Finally, in Table~\ref{t_sle_Q} we present the solution of systems of linear equations
with coefficients in $\mathbb{Q}$ (with arbitrary precision). The times in Sage are divided into
two different operations: the first one represents the time for obtaining a single solution,
and the second one the time for computing a basis of the null space of the original matrix.

\begin{table}
\centering
\begin{tabular}{| c | c | c | c |}
\hline
\multicolumn{1}{|c|}{{\bf Size (n)}} & \multicolumn{1}{c|}{{\bf Poly/ML}} & \multicolumn{1}{c|}{{\bf MLton}}
& \multicolumn{1}{c|}{{\bf Sage}} \\ \hline
 10 &   0.08 &   0.03 &  0.01 + 0.00 \\
 20 &   2.40 &   0.69 &  0.01 + 0.00 \\
 30 &  14.35 &   4.62 &  0.01 + 0.00 \\
 40 &  48.98 &  16.28 &  0.01 + 0.00 \\
 50 & 142.25 &  47.09 &  0.03 + 0.03 \\
 60 & 301.35 & 101.18 &  0.02 + 0.03 \\
 70 & 603.56 & 202.46 &  0.02 + 0.02 \\
 80 &    -   &    -   &  0.04 + 0.03 \\
 90 &    -   &    -   &  0.02 + 0.03 \\
100 &    -   &    -   &  0.03 + 0.02 \\
\hline
\end{tabular}
\caption{Elapsed time (in seconds) to compute the solution of a system of linear equations with coefficients in $\mathbb{Q}$.}
\label{t_sle_Q}
\end{table}

\section{Conclusions and Further work}
\label{s_conclusions_and_further_work}

Formalisation of Mathematics is a rather challenging, and sometimes
little rewarding, task. Formal proofs are considered deeply \emph{specific}
and remotely reusable, as well as very singular and concrete works.
As an example, the Isabelle/HOL Library offers,
to the best of our knowledge, three different representations of matrices,
each of them equally interesting and used for challenging works. Because of
this, the field has received little attention along the years, even inside
of the Formal Methods community (except for some groups of Computer Science
theorists, usually the ones involved in the development of the tools). Nevertheless,
the technology (both the hardware and the software) has improved along the last
decades at such rate that nowadays it is possible to face challenges that
were previously unthinkable.

Initiatives such as the Flyspeck project (lead by T.~Hales, for the formalisation
of the Kepler conjecture) and the classification of finite simple groups
(by G.~Gonthier) show that challenging results in Mathematics can be explored
with proving assistants. They also pave the way for research in some other fields
(for instance, algorithmics) where formalisation is also, at least, as relevant and required
as in Mathematics (see the previous failures mentioned above, that could be
critical in applications such as cryptology). In this field is where our
work can be considered as a first milestone in the way to the formalisation
of Linear Algebra algorithms; several of the tools that
have been presented and already formalised in our work are reusable
in Numerical Linear Algebra algorithms. For instance, some of
the serialisations introduced in Section~\ref{s_code_generation_to_functional_languages}
are already part of the Isabelle Library. Several proofs of basic properties
of elementary row and column operations are also reusable. Even with the simple
version of the Gauss-Jordan algorithm presented and formalised in this
work, the effort devoted and the results obtained pay off (formalisation
of previously unconsidered results, real world applications in digital
image processing, detection of commercial software bugs).

The amount of verified code generated in our work (\emph{ca.} $2\,500$ lines in each SML
and Haskell) is considerable and covers a wide range of applications in Linear Algebra.
Its formalisation (available in~\cite{ARDI13d}) took $15\,000$ lines of Isabelle code.
The HMA Library infrastructure reduced significatively the amount of mathematical
results to be formalised. A similar Library for Numerical Linear Algebra would help
in future developments.

As a natural continuation to our work, our intention is now to provide a framework
(and a methodology) in which the pieces presented here can be used to implement and
formalise Numerical Linear Algebra algorithms in a more \emph{general} way,
and which does not require a deep Isabelle knowledge. The ideas (and hopefully
part of the technology) are borrowed from the Autoref tool~\cite{LA13} (also developed
in Isabelle/HOL), which facilitates the refinement of data types (as we already did)
and, more interestingly, algorithms over abstract concepts (sets, maps) to algorithms
over concrete implementations, and automatically generates the refinement theorems
(that the user must prove). Consequently, one must prove the properties of the
original abstract algorithm (in which imperative constructs are also permitted),
and then ensure that the transformations (performed by monadic refinements)
preserve the original behaviour. Our original data types would be vectors
and matrices implemented as functions over finite domains (this work shows that
algorithmic formalisation is feasible within these types),
and algorithms over them, in a language closer to the one
presented in Algorithm~\ref{A:gauss} than to the Isabelle code snippets
presented. Input algorithms need not to be very optimised, to favour and
simplify their formalisation. Additionally, the HMA Library offers a background
to link the algorithms with their original mathematical signification.

In this setting, refined algorithms over optimised data structures
(such as \emph{iarrays}) become the concrete implementation, whose formalisation
shall emerge from iterated refinements of the original (abstract) algorithm,
paying attention exclusively to the particular transformations performed (and not
to the inherent complexities of the optimised version). The idea
has been successfully applied with some intricate algorithms in automata
theory~\cite{ESLANENISHSM13}. We aim at applying it to Numerical Linear Algebra
specifications of algorithms and refining them to compelling programs.

Incidentally, the methodology could also be applied to different fields
of Numerical Algebra where general simple algorithms are commonplace but
specialised versions (or refinements) are widely used by the community.

A different research line would be to explore \emph{certifying algorithms}.
Certifying algorithms permit to obviate the formalisation of the algorithm
specification, and focus on providing a certain output, and also a
\emph{certificate} (for instance, in the form of a witness) that the
output is \emph{correct} with respect to some requirements. Some of our
computations are apparently amenable to this kind of methodology (for
instance, computing the solutions to a system of linear equations and
the fundamental subspaces). On the other hand, certifying the computation
of the rank, determinant and even the rref of a matrix could be,
\emph{a priori}, as challenging as formalising them.


\section*{Acknowledgments}

David Matthews, Matthew Fluet and Tjark Weber gave us valuable advice
to speed up our programs in SML and Poly/ML. David Matthews also modified
Poly/ML 5.5.1 to improve the performance offered by variable binders.

Juan Luis Varona helped us to perform Mathematica\textsuperscript{\textregistered}
computing tests and gave us valuable information about this system. His work
encourages formalisation efforts and consequently motivates ours.

This work has been supported by the research grant FPI-UR-12, from Universidad de La Rioja.

\bibliographystyle{abbrv}

\begin{thebibliography}{15}

\bibitem{ALBOMERI}
    E.~Alkassar, S.~Böhme, K.~Mehlhorn and C.~Rizkallah.
    \newblock A Framework for the Verification of Certifying Computations.
    \newblock {\em J. Autom. Reasoning}, 2013. Accepted. Available from \url{http://dx.doi.org/10.1007/s10817-013-9289-2}.

\bibitem{ARDI13}
    J.~Aransay and J.~Divasón.
    \newblock Rank Nullity Theorem in Linear Algebra.
    \newblock Archive of Formal Proofs. 2013. \url{http://afp.sourceforge.net/entries/Rank_Nullity_Theorem.shtml}.

\bibitem{ARDI13b}
    J.~Aransay and J.~Divasón.
    \newblock Formalization and execution of Linear Algebra: from theorems to algorithms.
    \newblock In {\em G.~Gupta and R.~Peña eds., Pre-Proceedings of International Symposium on Logic-Based Program Synthesis and Transformation: LOPSTR 2013}: 49 -- 66. 2013.

\bibitem{ARDI13c}
    J.~Aransay and J.~Divasón.
    \newblock Performance Analysis of a Verified Linear Algebra Program in SML.
    \newblock In {\em L.~Fredlund and L.~M.~Castro eds. TPF 2013, V Taller de Programación Funcional}: 28 -- 35. 2013.
    \newblock Available from \url{http://babel.ls.fi.upm.es/prole2013/ActasPROLE_TPF.pdf}.

\bibitem{ARDI13d}
    J.~Aransay and J.~Divasón.
    \newblock {Gauss-Jordan elimination in Isabelle/HOL}.
    \newblock 2013. \url{http://www.unirioja.es/cu/jodivaso/Isabelle/Gauss-Jordan-2013-2/}.

\bibitem{DUPEVA13}
    A.~J.~Durán, M.~Pérez and J.~L.~Varona.
    \newblock Misfortunes of a mathematicians' trio using Computer Algebra Systems: Can we trust?
    \newblock {\em Submitted}, preprint available in \url{http://arxiv.org/abs/1312.3270}.

\bibitem{ESLANENISHSM13}
    J.~Esparza, P.~Lammich, R.~Neumann, T.~Nipkow, A.~Schimpf and J.~G.~Smaus.
    \newblock A Fully Verified Executable LTL Model Checker.
    \newblock In {\em N.~Sharygina and H.~Veith eds., Computer Aided Verification: CAV 2013, Lecture Notes in Computer Science} 8044: 463-478. Springer, 2013.

\bibitem{GO10}
    M.~S.~Gockenbach.
    \newblock Finite-dimensional Linear Algebra. CRC Press. 2010.

\bibitem{HA13}
    J.~Harrison.
    \newblock The HOL Light Theory of Euclidean Space.
    \newblock {\em J. Autom. Reasoning}, 50 (2): 173 -- 190. 2013.

\bibitem{HA13b}
    F.~Haftmann.
    \newblock Code generation from Isabelle/HOL theories.
    \newblock Technical Report, \url{http://isabelle.in.tum.de/dist/Isabelle2013-2/doc/codegen.pdf}.

\bibitem{HAKRKUNI13}
    F.~Haftmann, A.~Krauss, O.~Kuncar and T.~Nipkow.
    \newblock Data Refinement in Isabelle/HOL.
    \newblock In {\em S.~Blazy, C.~Paulin-Mohring and D.~Pichardie eds. Interactive Theorem Proving:
    4th International Conference Proving: ITP 2013, Lecture Notes in Computer Science Volume} 7998: 100-115. Springer, 2013.

\bibitem{HANI10}
    F.~Haftmann and T.~Nipkow.
    \newblock Code generation via higher-order rewrite systems.
    \newblock In {\em M.~Blume, N.~Kobayashi and G.~Vidal eds. Functional and Logic Programming: 10th International Symposium: FLOPS 2010, Lecture Notes in Computer Science}, 6009. Springer-Verlag, 2010.

\bibitem{HEDEMAMOPOSI12}
    J.~Heras, M.~Dénès, G.~Mata, A.~Mörtberg, M.~Poza and V. Siles.
    \newblock Towards a certified computation of homology groups for digital images.
    \newblock In {\em Computational Topology in Image Context: CTIC 2012, Lecture Note in Computer Science}, 7309: 49 -- 57. Springer, 2012.

\bibitem{LA13}
    P.~Lammich
    \newblock Automatic Data Refinement
    \newblock In {\em S.~Blazy, C.~Paulin-Mohring and D.~Pichardie eds. Interactive Theorem Proving: ITP 2013, Lecture Notes in Computer Science}: 7998: 84 -- 99. Springer, 2013.

\bibitem{NI11}
    T.~Nipkow.
    \newblock {Gauss-Jordan Elimination for Matrices Represented as Functions}.
    \newblock Archive of Formal Proofs, 2011. \url{http://afp.sourceforge.net/entries/Gauss-Jordan-Elim-Fun.shtml}.

\bibitem{NIPAWE02}
    T.~Nipkow, L.~C.~Paulson and M.~Wenzel.
    \newblock Isabelle/HOL: A Proof Assistant for Higher-Order Logic.
    \newblock In \emph{Lecture Notes in Computer Science}, 2283. Springer, 2002.

\bibitem{PA90}
    L.~C.~Paulson.
    \newblock Isabelle: The next 700 theorem provers.
    \newblock In \emph{P. Odifreddi, editor, Logic and Computer Science}, pages 361 -- 386. Academic Press, 1990.

\bibitem{RO08}
    S.~Roman.
    \newblock Advanced Linear Algebra (Third Edition). Springer. 2008.

\bibitem{ST13}
    C.~Sternagel.
    \newblock Proof Pearl - A Mechanized Proof of GHC's Mergesort.
    \newblock {\em J. Autom. Reasoning}, 51 (4): 357 -- 370. 2013.

\end{thebibliography}
%

\end{document}